\documentclass{ws-ijmpa}
\usepackage[super,compress]{cite}
\usepackage{graphicx}
\usepackage[colorlinks=true,linkcolor=blue,citecolor=blue,urlcolor=black]{hyperref}
\begin{document}
\markboth{Jonathan H. Davis}
{The Past and Future of Light Dark Matter Direct Detection}

%
\catchline{}{}{}{}{}
%

\title{The Past and Future of Light Dark Matter Direct Detection}

\author{JONATHAN H. DAVIS}
\address{Institut d'Astrophysique de Paris\\98 bis boulevard Arago\\75014 Paris\\France \\
jonathan.h.m.davis@gmail.com}

\maketitle

\begin{history}
\end{history}

\begin{abstract}
We review the status and future of direct searches for light dark matter. We start by answering the question: `Whatever happened to the light dark matter anomalies?' i.e. the fate of the potential dark matter signals observed by the CoGeNT, CRESST-II, CDMS-Si and DAMA/LIBRA experiments. We discuss how the excess events in the first two of these experiments have been explained by previously underestimated backgrounds. For DAMA we summarise the progress and future of mundane explanations for the annual modulation reported in its event rate.
Concerning the future of direct detection we focus on the irreducible background from solar neutrinos. We explain broadly how it will affect future searches and summarise efforts to mitigate its effects.

\end{abstract}

\ccode{PACS numbers: 95.35.+d, 95.85.Ry}

\tableofcontents

\section{Introduction}
In this review we summarise the progress which has been made by direct detection experiments in the past few years towards probing the interactions between light dark matter (DM) and nucleons, and the challenges faced by these experiments in the future. By `light' dark matter we refer to particles with masses $m$ in the approximate range $3\,\mathrm{GeV} \lesssim m \lesssim 20\,\mathrm{GeV}$.
Direct detection experiments aim to observe evidence of dark matter particles from the halo of our galaxy interacting with nuclei on Earth.
Signals of such scattering events correspond to $\sim$~keV-energy recoils of nuclei in these detectors. However there are many other potential `background' sources of nuclear recoils which could mimic a dark matter signal, and so these experiments seek to minimise these backgrounds as much as possible. For example the rates of cosmogenic events such as atmospheric muons are minimised by placing direct detection experiments deep underground and within layers of shielding. Radioactive decay of the materials surrounding the detector results in significant backgrounds, both from beta and gamma radiation inducing electronic recoils, and neutrons inducing nuclear recoils. As direct detection experiments get larger they will also have to contend with an irreducible background from neutrinos, particularly those from the Sun whose flux is large.
The observation of extra events in addition to those from expected background sources would in principle constitute a discovery of dark matter.
Hence the better these backgrounds are controlled and understood, the more sensitive the experiment will be to dark matter recoils. However conversely if backgrounds are underestimated this can lead to spurious claims of dark matter detection.

\begin{figure}[t]
\centering
\includegraphics[width=0.95\textwidth]{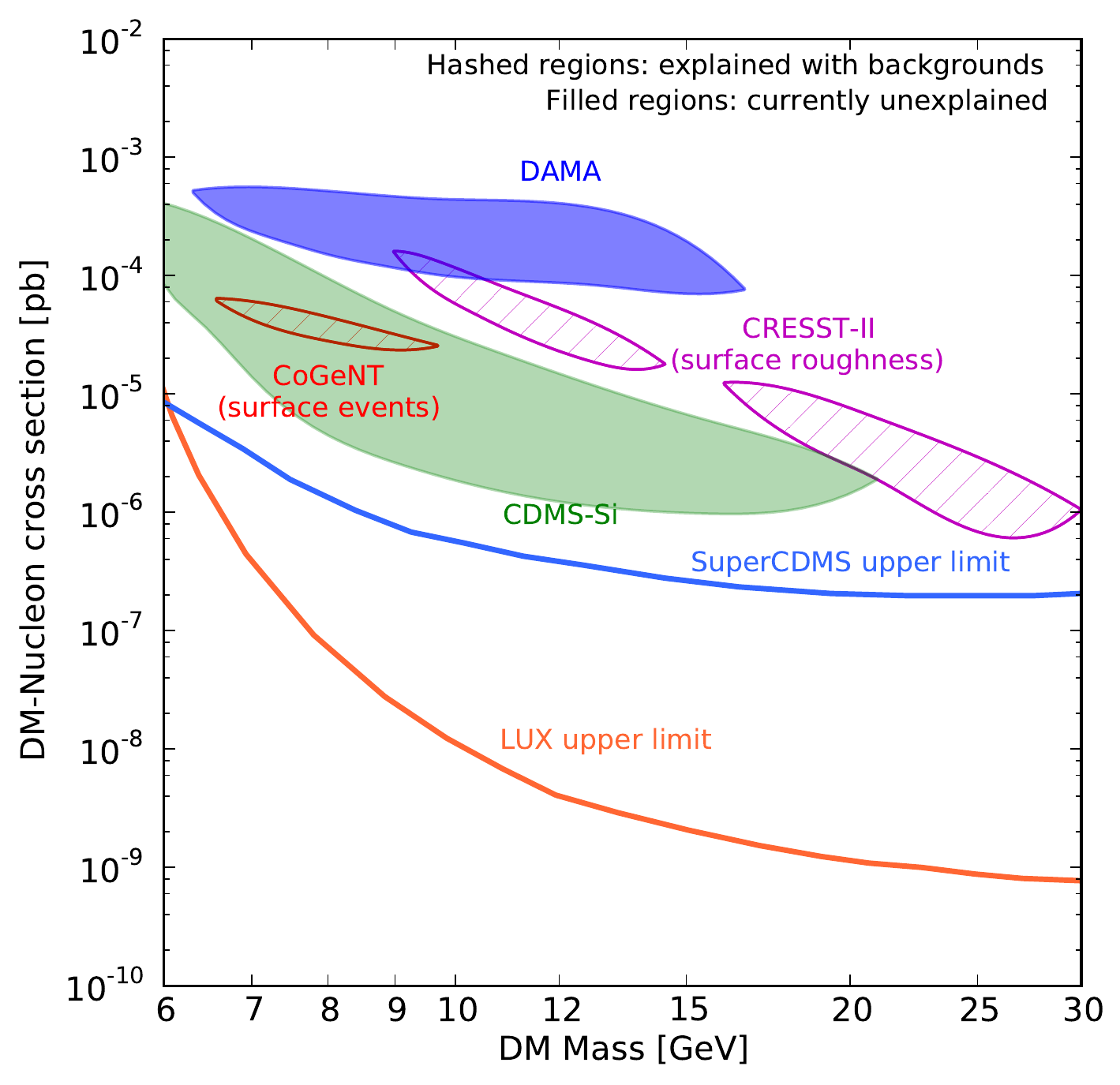}
\caption{Comparison of $90\%$ confidence upper limits on the DM-nucleon cross section set by the LUX~\cite{Akerib:2013tjd} and SuperCDMS~\cite{Agnese:2014aze} experiments with preferred regions (at various levels of statistical significance) claimed by the CoGeNT~\cite{Aalseth:2012if}, CRESST-II~\cite{Angloher:2011uu}, DAMA/LIBRA~\cite{Bernabei:2013xsa,Bernabei:2008yi,Bernabei:2013cfa} and CDMS-Si~\cite{Agnese:2013rvf} experiments for positive signals of dark matter recoils. Hashed regions for the CoGeNT and CRESST-II experiments indicate that the excess events in these detectors, previously taken as tentative evidence for dark matter, have now been explained entirely by underestimated backgrounds (indicated in parentheses)~\cite{Davis:2014bla,Angloher:2014myn,Kuzniak:2012zm}. The filled regions represent anomalies which may have a dark matter or mundane explanation.}
\label{fig:dm_exclusion}
\end{figure}

The experimental status of light dark matter direct detection in past few years has been dominated by competition between experiments claiming tentative signals of dark matter discovery, and those setting exclusion limits due to null results. However more recently it has become increasingly clear that the null scenario is the more likely, the reasons for which we will elucidate in this review.
These potential signals are referred to generically as `anomalies', in the form of excess events seen by the CoGeNT~\cite{Aalseth:2012if}, CRESST-II~\cite{Angloher:2011uu,Angloher:2014myn} and CDMS-Si~\cite{Agnese:2013rvf} experiments above their expected backgrounds. Additionally there is the DAMA/LIBRA collaboration~\cite{Bernabei:2013xsa,Bernabei:2008yi,Bernabei:2013cfa}, who have been claiming discovery of dark matter for over a decade due to an annual variation seen in their observed events. 

The interest surrounding these anomalies stems from the fact that they could all be interpreted in terms of dark matter interacting with nuclei, leading to preferred regions such as shown in figure~\ref{fig:dm_exclusion}. However null results from other experiments, particularly XENON100~\cite{Aprile:2012nq} (but also e.g. XENON10~\cite{Angle:2011th}, EDELWEISS~\cite{Armengaud:2012pfa}, ZEPLIN-III~\cite{Akimov:2011tj} and CDMS-II~\cite{Ahmed:2009zw}), placed strong constraints on dark matter interpretations of these anomalies. Negative results from the LUX~\cite{Akerib:2013tjd} and SuperCDMS~\cite{Agnese:2014aze} experiments followed (and also e.g. CDEX~\cite{Yue:2014qdu} and CDMSlite~\cite{Agnese:2013jaa}), which made it impossible to explain these anomalies with dark matter elastically scattering with nuclei while evading upper bounds from all null searches, as can be seen in figure~\ref{fig:dm_exclusion}.
Hence there has been tension between two sets of experiments: those claiming discovery signals and those setting upper limits. Solutions to this tension fall into two main categories:
\begin{enumerate}
\item The first is to change how dark matter particles interact with nuclei.
Indeed their interactions with nuclei are unknown and may be more complicated than simple elastic scattering, which can weaken the upper limits set by null results with respect to the preferred regions from positive searches~\cite{Frandsen:2013cna,Chang:2008gd,Chang:2010yk,TuckerSmith:2001hy}. Such explanations appeared initially promising, however due to the strength of bounds from LUX, SuperCDMS and other experiments most of these models are now still in some tension~\cite{Chen:2014tka}.
\item The second is that some or all of the positive results are not due to dark matter particles, but to underestimated backgrounds. This is particularly relevant for light dark matter particles as they should result in nuclear recoils with energies just above the typical threshold of a direct detection experiments, where the backgrounds are less well understood.
 Indeed much of the tension has now been resolved due to precisely this reason, as we discuss in section~\ref{sec:past} with respect to the CRESST-II~\cite{Angloher:2011uu,Angloher:2014myn} and CoGeNT~\cite{Aalseth:2012if} experiments.
\end{enumerate}

Much of this review focuses on the fate of the `anomalies' observed by the CoGeNT, CRESST-II and DAMA/LIBRA experiments in the context of this second option. 
In section~\ref{sec:past} we discuss the explanations of the first two anomalies in terms of underestimated backgrounds, while section~\ref{sec:dama} deals with progress and challenges in the explanation of the DAMA/LIBRA data both with dark matter and mundane sources. We also discuss in section~\ref{sec:nu_bg} the future challenges which all direct detection experiments will face in separating any potential light dark matter signals from the potentially large solar neutrino background. We present a short summary of the theory behind dark matter direct detection in~\ref{sec:app1}, however the interested reader should consult other reviews such as refs.~\citen{Cerdeno:2010jj,Freese:2012xd,Schumann:2015wfa,Bertone:2004pz} for more information.

\section{Light dark matter anomalies in CoGeNT and CRESST-II explained with backgrounds \label{sec:past}}
In this section we review the fate of the anomalous events observed by the CoGeNT~\cite{Aalseth:2012if} and CRESST-II~\cite{Angloher:2011uu} experiments. In both cases these anomalies took the form of a large number of additional low-energy recoils, which could not (at the time) be accounted for by the known backgrounds from e.g. natural and cosmogenic radioactivity in materials surrounding the detector. 
These were initially taken as evidence for dark matter interactions primarily due to their spectra,  however they have both now been explained by underestimated backgrounds.

\subsection{The CoGeNT excess from surface events}

Here we summarise the analysis performed by the CoGeNT collaboration which led to an erroneous preference for dark matter recoils in their data, represented by the `region of interest' in figure~\ref{fig:dm_exclusion}.
The CoGeNT experiment\cite{Aalseth:2012if,Aalseth:2014eft} works using a p-type point-contact germanium detector. An event in the detector constitutes a change in the measured voltage over a time of a few microseconds. For each event the magnitude of this change is proportional to the recoil energy, while the duration is measured as the rise-time.  
The majority of the active volume of the CoGeNT detector is a p-type semiconductor with a charge collection efficiency $\epsilon$ of unity, referred to as the bulk of the detector. 
Towards the outer edge of the detector modules is the millimetre-thick transition layer where $0 < \epsilon < 1$. Events occurring here are denoted as surface events\cite{Aalseth:2012if,Aalseth:2014eft}. 

Backgrounds (e.g. from radioactivity) induce events preferentially towards the outside of the detector volume. However dark matter particles are weakly-interacting and so make no distinction between the surface and bulk of the detector.
As such they are considerably more likely to scatter in the bulk than the surface, whose volume is much smaller. Hence the surface population is dominated by background events and as such needs to be removed or accounted for before an analysis of the CoGeNT data-set for dark matter recoils. Additionally the partial charge collection means these surface events will be measured with typically lower energies, resulting in a spectrum with a low-energy rise which can mimic the recoil spectrum of light dark matter. This makes surface events particularly dangerous as a background.

\begin{figure}[t]
\centering
\includegraphics[width=0.99\textwidth]{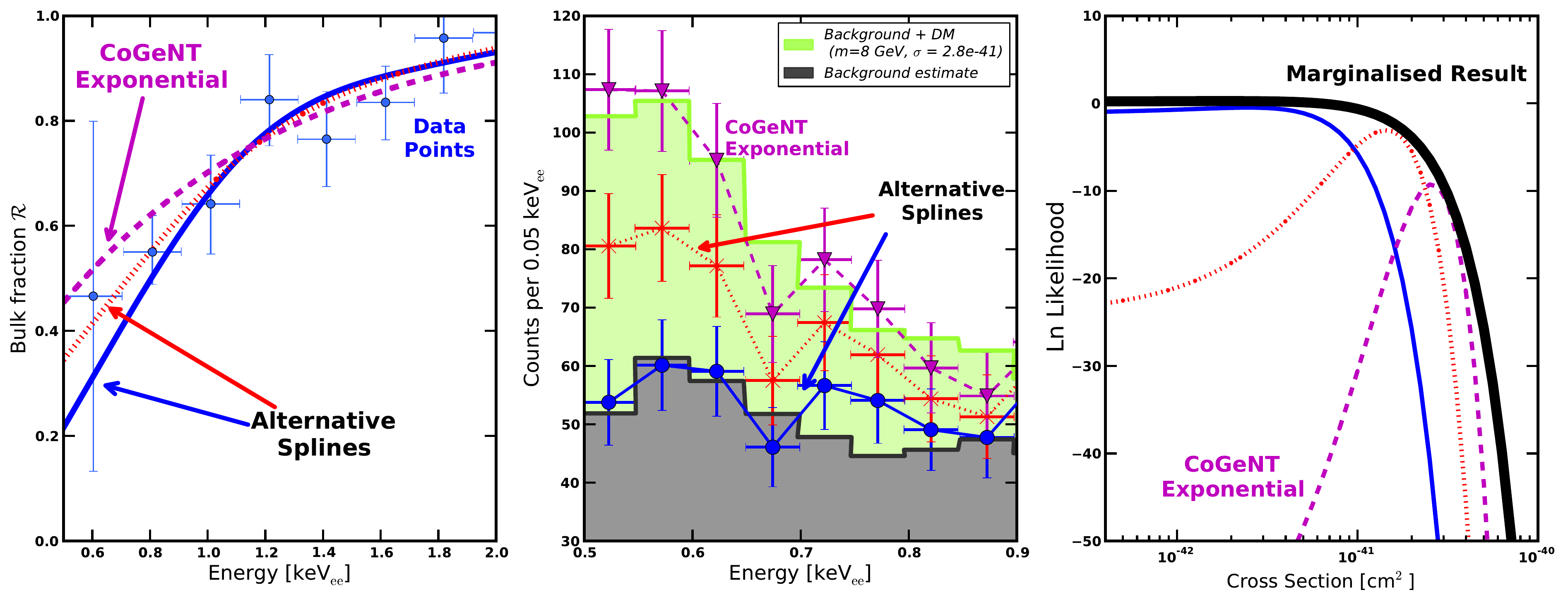}
\caption{The effect on a search for dark matter particles with a mass of 8~GeV due to uncertainties in the relative sizes of the bulk and surface event populations in the CoGeNT data-set~\cite{Aprile:2012nq}, where the latter population is dominated by background events. Different parameterisations for the bulk fraction $\mathcal{R}$ (right panel) lead to different spectra for the bulk events (central panel) and therefore different levels of preference for a dark matter signal (right panel). The CoGeNT collaboration used only the function labelled as `CoGeNT exponential' leading to a strong preference for dark matter recoils, however when marginalising over the full uncertainty in the choice of function for $\mathcal{R}$ the preference is less than $1 \sigma$~\cite{Davis:2014bla}.
This figure has been reproduced under a  Creative Commons Attribution 3.0 License from ref.~\citen{Davis:2014bla}.}
\label{fig:cogent}
\end{figure}

In order to separate the bulk and surface event populations the CoGeNT collaboration employed the following method in ref.~\citen{Aalseth:2012if}. Full details of the analysis can be found in refs.~\citen{Davis:2014bla,Aalseth:2012if}.
The spectrum of the full CoGeNT data-set was converted to that of pure bulk events (which may contain a dark matter signal or only background events) by multiplication by an energy-dependent function called the bulk fraction $\mathcal{R}$. For an energy $E$ this gives the fraction of the total number of events with energies between $E - \Delta E / 2$ and $E + \Delta E / 2$ which occur within the bulk of the detector, where $E$ is the central energy value and $\Delta E$ is the bin size. 
The bulk and surface events are separated using their rise-times, with the surface events being slower on average, leading to longer rise-times.
However on an event-by-event basis one does not know whether an energy-deposit occurred in the bulk or on the surface, and so $\mathcal{R}$ was determined by the CoGeNT collaboration for different energy bins by statistical methods. Specifically the distributions of bulk and surface events in rise-time were fit with two separate log-normal distributions (other distributions have also been considered~\cite{Davis:2014bla}). This six-parameter fit leads to significant uncertainties on the form of $\mathcal{R}$.

Values of the bulk fraction $\mathcal{R}$ as determined by the CoGeNT collaboration for their $807$ live days data-set are shown as data-points on the left panel of figure~\ref{fig:cogent}. The large error bars arise from the uncertainties in the log-normal fits to the rise-times and are particularly large at low energies where the bulk and surface populations are more difficult to separate. The spectral shape which the data-points follow is partially due to energy-dependent cuts on the rise-time performed by the CoGeNT collaboration. The CoGeNT collaboration fit this data with a one-parameter exponential function of form $\mathcal{R}(E)_{\mathrm{CoGeNT}} = 1  - \mathrm{exp}(- \alpha E)$, with the constant $\alpha$ determined from this fit to be $\alpha = 1.21 \pm 0.11$. This is labelled as `CoGeNT exponential' in the left-panel of figure~\ref{fig:cogent} for $\alpha = 1.21$.

By multiplying the pure CoGeNT spectrum with  $\mathcal{R}(E)_{\mathrm{CoGeNT}}$ (and applying corrections for the detector efficiency and known excitation peaks etc.) the collaboration obtained the spectrum of events labelled in the central panel of figure~\ref{fig:cogent} as `CoGeNT exponential'. This was obtained using $\alpha = 1.21$, though either $\alpha = 1.10$ or $\alpha = 1.32$ give very similar spectra. If  $\mathcal{R}(E)_{\mathrm{CoGeNT}}$ represents the correct bulk fraction then this spectrum should be that of purely bulk events. As shown in the central panel of figure~\ref{fig:cogent} when using this function for the bulk fraction there is a significant excess of low-energy events above the expected background in the bulk which fits well to the expected spectrum from light dark matter recoils. Indeed the likelihood function shown in the right panel of figure~\ref{fig:cogent} shows a clear preference for cross sections just above $10^{-41}$~cm$^2$ for dark matter with a mass of 8~GeV i.e. the centre of the region of interest corresponding to the CoGeNT anomaly.

However this is not the only possible choice for $\mathcal{R}(E)$. As shown in the left panel of figure~\ref{fig:cogent} there are a wide-range of functions which fit well to the data for the bulk fraction within the large error bars. Since there is no theoretical motivation for choosing $\mathcal{R}(E)_{\mathrm{CoGeNT}} = 1  - \mathrm{exp}(- \alpha E)$ any function which fits the data for $\mathcal{R}$ is equally acceptable \emph{a priori}. However the CoGeNT collaboration did not publish results using any other function for $\mathcal{R}(E)$. Indeed if we instead use cubic splines for the form of $\mathcal{R}(E)$ as labelled in the left panel of figure~\ref{fig:cogent} then it is clear from the central panel that the spectrum of bulk events looks radically different, compared to the version derived by the CoGeNT collaboration. Indeed for the solid blue spline the data are consistent with the expected background. 

Hence the uncertainties in the amount of contamination from surface events are much larger than assumed by the CoGeNT collaboration through their assumption that the bulk fraction took the form of $\mathcal{R}(E)_{\mathrm{CoGeNT}} = 1  - \mathrm{exp}(- \alpha E)$. When the full uncertainty in the functional form of $\mathcal{R}(E)$ was taken account of in ref.~\citen{Davis:2014bla}, through marginalising over all functions, it was found that there was less than one-sigma evidence for dark matter in the CoGeNT data-set. Indeed any preference for dark matter was a result of a bias in the analysis from an underestimation of the uncertainties in the ratio of surface to bulk events. Said differently: the low-energy rise typical of surface events mimicked a low-mass dark matter signal, as seen in the central panel of figure~\ref{fig:cogent} for the `CoGeNT exponential'.

This result has been confirmed with the more recent 1129 live days data-set from CoGeNT~\cite{Aalseth:2014eft} in analyses by refs.~\citen{Davis:2014bla} and \citen{Aalseth:2014jpa} and also by similar experiments such as CDEX~\cite{Yue:2014qdu}. It is clear that the CoGeNT data is now fully consistent with known backgrounds from the bulk and surface of the detector, and so there is no preference for a dark matter recoil signal. The issue of the CoGeNT excess highlights the dangers of underestimated backgrounds for direct detection experiments, especially those which mimic the low-energy rise expected from light dark matter.

\subsection{Surface roughness and CRESST-II}
In 2011 a group within the CRESST-II (Cryogenic Rare Event Search with Superconducting
Thermometers) collaboration claimed evidence for a population of low-energy nuclear recoils in their detector which could not be explained by any known background source~\cite{Angloher:2011uu}. This low-energy rise was similar to that observed by the CoGeNT experiment and so could be interpreted broadly with the same dark matter mass and cross section (although the agreement was not perfect, as shown in figure~\ref{fig:dm_exclusion}).  Additionally this result was potentially more credible since the CRESST-II experiment has the ability to distinguish nuclear and electronic recoils. Indeed the CRESST-II experiment is constructed from scintillating CaWO$_4$ crystals, for which recoil events generate signals in both phonons and scintillation light. The ratio of these two signals determines which type of recoil the event originates from~\cite{Angloher:2011uu,Angloher:2014myn}.

However the authors of ref.~\citen{Kuzniak:2012zm} showed that the study which lead to this claim of a tentative dark matter signal~\cite{Angloher:2011uu} underestimated a particular background originating from the interactions of $^{206}$Pb nuclei (which themselves originate from the decay of $^{210}$Po) on the surface of the clamps holding the detector in place. The interactions of these $^{206}$Pb nuclei near the surface of the clams generated cascades of secondary recoils in the CRESST-II detector. These were modelled in ref.~\citen{Angloher:2011uu}, however the authors assumed that the surface of the clamps was perfectly smooth. When the more realistic scenario of a rough surface for the clamps was taken into account this background was shifted partially to lower energies, as some of the energy from these cascades was absorbed by the rough surface instead of reaching the detector~\cite{Kuzniak:2012zm}.
This lead to a low-energy rise in the spectrum of nuclear recoil events from this background~\cite{Kuzniak:2012zm}, giving a good fit to the `excess' of events in the study of ref.~\citen{Angloher:2011uu} which was originally taken as tentative evidence for dark matter recoils.

Indeed a more recent study by the CRESST-II collaboration~\cite{Angloher:2014myn} using a new set-up without these troublesome clamps found no evidence for dark matter recoils. Hence the collaboration set an upper limit on the DM-nucleon cross section which excluded the preferred region found in ref.~\citen{Angloher:2011uu}. As with the case of CoGeNT this demonstrates the extreme difficulty in modelling backgrounds for direct searches, and the limitations of simulations in correctly capturing the physics of the real system.

\section{Towards an understanding the origin or the DAMA annual modulation \label{sec:dama}}
The DAMA/LIBRA and the former DAMA/NaI experiment (referred to hereafter together as DAMA) have accumulated over ten years worth of data in the search for dark matter recoils~\cite{Bernabei:2013xsa,Bernabei:2008yi,Bernabei:2013cfa}. The DAMA apparatus is composed of approximately 250~kg of NaI target surrounded by layers of shielding and located underground at the Gran Sasso lab~\cite{Bernabei:2013xsa,Bernabei:2008yi,Bernabei:2013cfa}. The DAMA collaboration currently claim to have observed an annual modulation in their data at a level of around $2\%$ (the modulation fraction) of the total event rate and with a statistical significance of $9.2\sigma$. This modulation is approximately sinusoidal and peaks in late May. Additionally the collaboration claim to observe a modulation only at low energies, below approximately 6~keV$_{\mathrm{ee}}$, and only in the single-hit event population (i.e. events which occur alone within a small time-interval, in contrast to multiple-hit events). 

Due to the Earth's orbit around the Sun the relative velocity between dark matter particles in the galactic halo and terrestrial direct detection experiments varies over the year, leading to an approximately sinusoidal modulation for the recoil rate of DM particles at keV-energies which peaks in early June (see \ref{sec:app1} and also e.g. refs.~\citen{Freese:2012xd,McCabe:2013kea}).
Hence due to its similarity with the annual modulation expected from a dark matter recoil signal  the DAMA collaboration have claimed that their data is evidence for dark matter particles scattering in their detector. However the most basic scenario of dark matter scattering elastically (and coherently) with nuclei requires a cross section which is excluded by other direct detection experiments such as LUX~\cite{Akerib:2013tjd} and SuperCDMS~\cite{Agnese:2014aze}. This is because the rate of modulated events observed by DAMA is approximately 100 counts per day, and so a large interaction cross section is needed to provide this many events (there are also issues with the unmodulated event rate, which we discuss later).
Hence in this section we start by discussing the challenges of explaining the DAMA signal with dark matter, before following with a discussion of models for the DAMA modulation which employ only Standard Model particles.

\subsection{Challenges for dark matter}
A popular method of alleviating the tension between a dark matter explanation for the DAMA signal and the exclusion limits from null experiments is to alter how the dark matter interacts with nuclei.
The standard assumption is that dark matter particles scatter elastically and coherently with nuclei. `Elastic' implies that the dark matter particle does not change state upon scattering, however it could instead transition to either a heavier state (inelastic) or a lighter state (exothermic)~\cite{Frandsen:2013cna,Chang:2008gd,Chang:2010yk,TuckerSmith:2001hy}. In this case the recoil spectrum is altered in a way which depends upon the mass of the target nucleus, allowing the rate at DAMA to be enhanced relative to the xenon-based experiments (e.g. LUX and XENON100), which set the strongest exclusion limits. 

Coherent scattering means that the dark matter scatters identically with protons and neutrons, resulting in the scattering rate scaling with the square of the atomic number (see~\ref{sec:app1}). If this assumption is relaxed then the proton and neutron terms can be made to interfere destructively for e.g. xenon-based targets only, thereby weakening the limits on the DAMA region.
Alternatively the scattering rate can also be altered such that it depends on the velocity of the dark matter, due to the dark matter interacting with nucleons via a pseudo-vector coupling~\cite{Dienes:2013xya}. Since the dark matter velocity is non-relativistic this coupling is therefore heavily suppressed and again depends on the mass of the target nucleus.

However most of these models still find it difficult to evade all of the constraints from e.g. LUX~\cite{Akerib:2013tjd} and SuperCDMS~\cite{Agnese:2014aze}, especially since the latter uses a germanium target which is not as heavy as xenon, and also those from e.g. flavour experiments~\cite{Dolan:2014ska}. For the case of spin-dependent scattering (see~\ref{sec:app1}) constraints from the PICO~\cite{Amole:2015lsj} and KIMS~\cite{Lee.:2007qn} experiments are also difficult to evade~\cite{DelNobile:2015lxa}.
Hence models which explain DAMA while evading all other constraints face the problem of being strongly fine-tuned or else of becoming increasingly complex e.g. composite models of dark matter~\cite{Wallemacq:2014sta} or mirror matter~\cite{Foot:2014xwa}.

A dark matter explanation (and indeed any model invoking a new population of events) faces an additional obstacle: the unmodulated spectrum of events in DAMA is well-fit by known radioactive backgrounds~\cite{Pradler:2012qt,Pradler:2012bf}. This means that even at low-energy (where the annual modulation is observed) there is little room for extra events in the data above background and so any new population will need a large modulation fraction to explain the DAMA signal i.e. if the DAMA data shows an $\sim 2\%$ modulation and the dark matter forms only $10\%$ or less of the unmodulated signal then the dark matter itself needs a $\sim 20\%$ modulation fraction or larger. So even if a dark matter signal could provide the required number of modulated events, if its modulation fraction is too small then it will over-shoot the unmodulated spectrum.

The modulation fraction from dark matter depends to some extent on the distribution of dark matter velocities in the galactic halo $f(v)$ (see~\ref{sec:app1}), which is \emph{a priori} unknown. Generally this distribution is assumed to take the form of a Maxwell-Boltzmann distribution with a velocity dispersion of $\sigma_v = \sqrt{3/2} v_0$ where $v_0 = 220$kms$^{-1}$, cut off at the escape velocity. This is called the Standard Halo Model (SHM). In this case the modulation fraction is expected to be of a few percent in size.
 However there are many other forms which this distribution can take, for example those motivated by numerical simulations such as in refs.~\citen{Mao:2013nda,McCabe:2010zh} or with streams of dark matter~\cite{Savage:2006qr}. 

It has been suggested that using one of these alternative forms for $f(v)$ could result in a different (and potentially larger) modulation fraction~\cite{Freese:2012xd}, thereby making a dark matter scenario more favourable for the DAMA data.
However many authors have developed techniques to analyse the DAMA data and how it compares to null results from other experiments without having to make assumptions about the astrophysical velocity distribution~\cite{Fox:2010bz,McCabe:2011sr,Anderson:2015xaa,Feldstein:2014ufa}. Such velocity-independent methods have shown that it is very difficult to explain DAMA with dark matter for any choice of halo function, however arbitrary. Indeed the DAMA modulation is even inconsistent with its own unmodulated event rate for any velocity distribution~\cite{HerreroGarcia:2012fu}. Hence although the possibility to explain the DAMA data with dark matter may remain, such models are no longer the most simple option.

\subsection{Progress for non-DM models of the DAMA events}
Dark matter recoils are not the only method of explaining the DAMA signal.
Models which explain the DAMA modulation without dark matter fall into two main categories: 
\begin{enumerate}
\item A source of modulated events on top of the known un-modulated backgrounds. In this respect these models are similar to those which invoke dark matter, however they generally rely on cosmogenic particles such as muons or neutrinos to provide the modulation instead~\cite{Davis:2014cja,Blum:2011jf,Ralston:2010bd}.
\item The modulation is explained as an artefact of the detector itself or an error introduced when the raw data is processed. In this case there is no modulated population of events, but instead the un-modulated event rate appears to modulate artificially. 
\end{enumerate}

For option (1) it is possible to get the right phase of the modulation and make strong predictions for other experiments, however such models face the same difficulties regarding the unmodulated spectrum of events as dark matter. The mechanism for generating these modulated events also needs to be unique to DAMA, in order to get past the null results from other modulation searches~\cite{2012arXiv1203.1309C}.
Models which fall into this category are rather few compared to the number of dark matter explanations for the DAMA signal. They generally have the events seen by DAMA generated in some way by atmospheric muons alone~\cite{Blum:2011jf,Ralston:2010bd} or in combination with solar neutrinos~\cite{Davis:2014cja}, since the rates of these events are known to modulate with an annual period. Indeed the production of atmospheric muons by decaying cosmic ray particles in the stratosphere is correlated with temperature, with a maximum for the annual mode approximately 30~days later than the observed DAMA modulation~\cite{Ambrosio:1997tc,FernandezMartinez:2012wd,D'Angelo:2011fs}. The muon signal also possesses significant power at periods longer than one year, including an approximately $1\%$ modulation at a period of 11~years from solar activity~\cite{Ambrosio:1997tc,FernandezMartinez:2012wd,D'Angelo:2011fs}. Solar neutrinos also modulate with a period of one year due to the changing distance between the Earth and Sun, and so their flux is largest around January 4th~\cite{Bellini:2013lnn,PhysRevD.78.032002}.
Alone the atmospheric muon and solar neutrino modulations have the wrong phase to fit the DAMA signal, however in combination they can partially interfere to give the required phase, period and modulation fraction~\cite{Davis:2014cja}. Hence based purely on timing information the fit of such cosmogenic models to the DAMA data is as good as from dark matter, and so the closeness of the phase of the DAMA data to that expected from the dark matter rate is not a `smoking-gun' for the latter.

These cosmogenic sources may not directly be responsible for the DAMA modulated events, but likely through a secondary such as neutrons, which are liberated from material surrounding the detector by the muons and neutrinos~\cite{Blum:2011jf,Ralston:2010bd,Davis:2014cja}. The spectrum and rate of the produced neutrons can depend strongly on the detector environment, and so could be larger at DAMA than other experiments such as LUX. However the exact mechanism to generate the DAMA events from cosmogenic sources is unknown, and current simulations show that it may be difficult to produce enough neutrons in this way~\cite{Klinger:2015vga,Jeon:2015sya}, at least if the neutrons are scattering elastically in the detector. Hence for now the actual rate of events which could be generated by these cosmogenic sources is unknown. Indeed as for dark matter recoils any new source must also fit well to the unmodulated DAMA spectrum, which likely means that it must have a large modulation fraction or that some of the backgrounds at low-energy experience a modulation themselves. 

An important point for such models is that if the rate at which neutrons, or any alternative secondary, are generated from muons or neutrinos is non-linear then their modulation fraction could be much larger than for the individual cosmogenic sources themselves (which is around a few percent). For example if the secondary rate $R_2$ (e.g. neutrons) depends on the primary rate $R_1$ (e.g. muons or neutrinos) as $R_2 \sim R_1^{\alpha}$ then the modulation residual $\Delta R_2 / R_2 \approx \alpha \Delta R_1 / R_1$.
Hence until we know all of the methods by which neutrons (or other secondaries which could lead to the DAMA events) are generated from cosmogenic sources we will not know their modulation fraction precisely, and so it is difficult to rule them out based on the modulation fraction of the primary source.
Alternatively the mechanism for generating the DAMA events may not involve nuclear recoils. Since the DAMA detector can not distinguish whether an event is due to a nuclear recoil or some other source (e.g. a photon) it is possible even that photon emission induced by electron capture, responsible for a large population of events in the DAMA detector at low-energy~\cite{Pradler:2012qt,Pradler:2012bf}, could be induced to modulate by cosmogenic sources and so explain the DAMA events. However this is purely speculative at this stage and it is clear that there is much scope for an improved understanding of annually modulated sources for the DAMA events.

Regarding option (2) this is the most attractive explanation from the perspective of the spectrum of the unmodulated events seen by DAMA, since it does not require any additional population of events. However it is not clear why the phase of any artificial effect would be close to that seen in the DAMA modulation, and why this would not also affect the multiple-hit event population (for example). Additionally it is harder to use such a model to make predictions for other direct detection experiments looking to replicated the DAMA signal, though to some extent any null result lends credence to this claim~\cite{2012arXiv1203.1309C}. 

In all of the above cases: either dark matter, a new mundane event source or a detector effect, new results from other DAMA-like experiments will be highly beneficial. Examples of such experiments are DM-Ice~\cite{Cherwinka:2014xta}, ANAIS~\cite{Amare:2013lca}, KIMS~\cite{kims_pres} and SABRE~\cite{sabre_pres}. We summarise the different potential results at a second DAMA-like experiment in a different lab and its implications for DAMA below:
\begin{itemize}
\item \emph{A second experiment observes a modulation with the same phase and modulation amplitude as DAMA.} This would constitute strong evidence for a dark matter interpretation of the DAMA signal, especially if the second experiment is in the southern hemisphere where effects correlated with the seasons (e.g. the modulation of atmospheric muons) have their phases shifted by half a year. 
\item \emph{A second experiment observes a modulation with a different phase from DAMA.} Such a result would imply that the DAMA modulation is due to an additional source of mundane events i.e. not dark matter. In this case the phase of this second signal combined with that from DAMA could be used to pin down the exact model, especially if it is due to an interference effect between two sources~\cite{Davis:2014cja}.
\item \emph{A second experiment observes no modulation in its event rate.} In this case the DAMA modulation is likely due to a detector effect (i.e. option (2) above) rather than an extra source of events. There is already a null result from a modulation search by CDMS~\cite{2012arXiv1203.1309C}, however a definitive result can only really come from an experiment constructed from NaI (like DAMA) for which the interactions between dark matter and nuclei in the detector should be identical.
\end{itemize}

\section{Solar neutrino backgrounds for light dark matter searches \label{sec:nu_bg}}
\begin{figure}[b]
\centering
\includegraphics[width=0.9\textwidth]{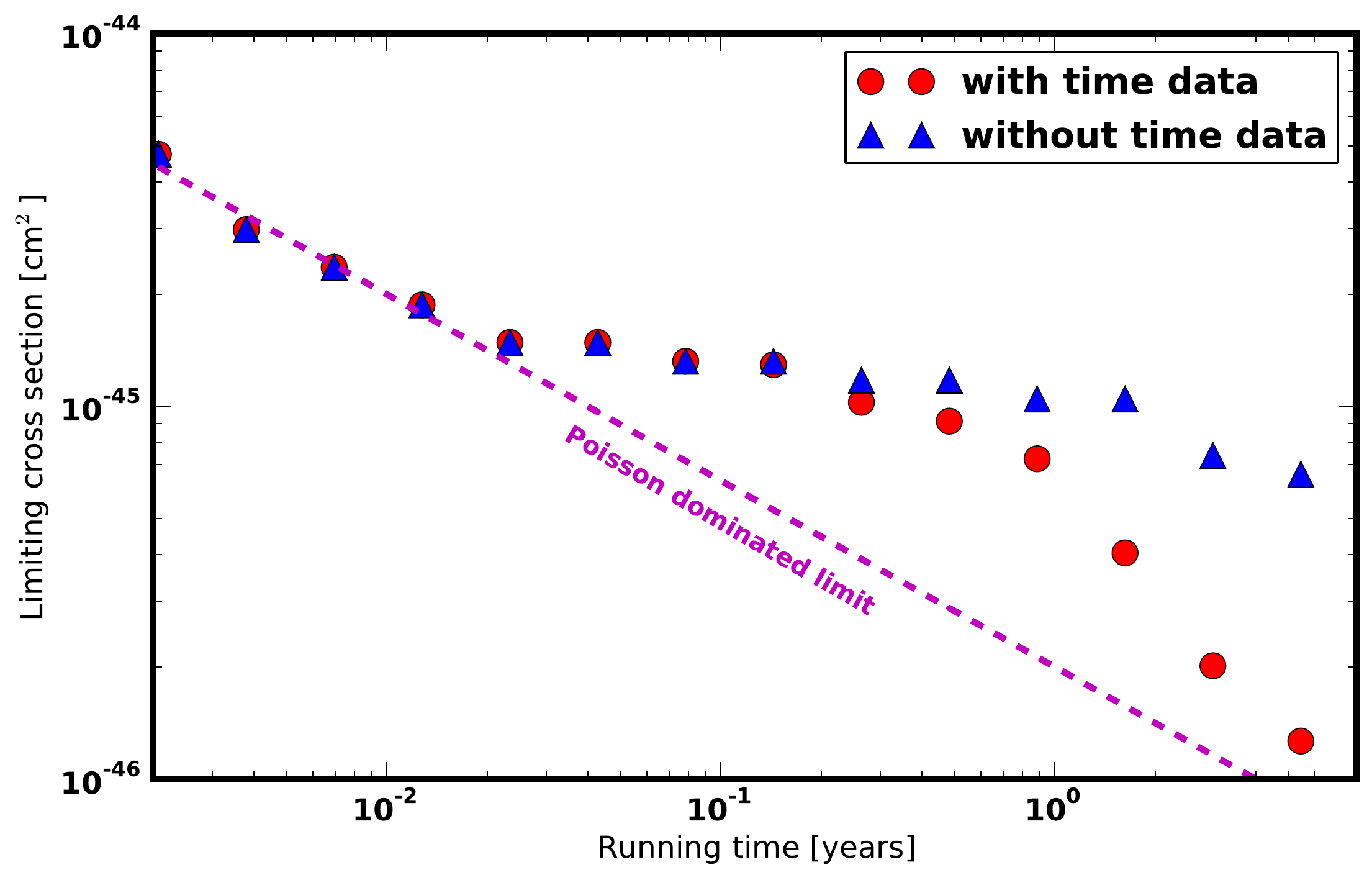}
\caption{Values of the $90\%$ upper limit for a ten tonne low-threshold xenon experiment as a function of running time, and for a dark matter mass of 6~GeV. Initially the limiting cross section scales as the inverse square root of running time, as expected from Poisson statistics.
Using only information on the spectrum of events the limiting cross section reaches a saturation value where it remains mostly constant with increasing running time. However with the different temporal variation of the solar neutrino and dark matter signals used as an additional discrimination parameter the upper limit returns to the Poisson-dominated case, once enough statistics have been accumulated.
This figure has been reproduced under a  Creative Commons Attribution 3.0 License from ref.~\citen{Davis:2014ama}.}
\label{fig:lims_nu}
\end{figure}

We have seen that a full understanding of backgrounds is crucial for a direct search for light dark matter, especially if these backgrounds have a similar spectrum to that expected from dark matter recoils. For the current generation of experiments all such backgrounds are in principle `reducible' i.e. they can be brought to a nominal level using shielding, for example. One such example is natural radioactivity of the material surrounding the detector. This can be reduced through techniques such as volume fiducialisation for xenon experiments~\cite{Akerib:2013tjd,Aprile:2011dd}, which employ the self-shielding properties of xenon by using only the innermost volume of the detector  for a dark matter search.  
If the number of background events remains constant for increasingly large detectors then their  limiting sensitivity scales linearly with exposure (which is running time multiplied by fiducial volume) i.e. $\sigma_{\mathrm{lim}} \sim (\mathrm{exposure})^{-1}$. Indeed most projections for future experiments assume `zero-background' which is in principle attainable provided all of the background events arise from reducible sources.

However future multi-tonne experiments will be large enough such that solar neutrinos will lead to non-negligible numbers of nuclear-recoil events in these detectors, with recoil energies of a few keV. Since neutrinos are weakly interacting they can not be shielded against and so constitute an `irreducible' background. Additionally their spectrum is almost identical to that expected from light dark matter~\cite{Billard:2013qya,Davis:2014ama}, which makes such a background particularly troublesome. The DM-nucleon cross sections at which this neutrino background is of similar size to (or larger than) the dark matter recoil signal is referred to as the `neutrino floor'.
Despite its name the neutrino floor does not represent an absolute limit on the sensitivity of direct detection experiments. Instead it represents broadly two effects: The \emph{first and most fundamental} is that the sensitivity of future experiments is limited fundamentally by Poisson uncertainties on the size of the neutrino background i.e. if a DM signal is smaller than these uncertainties then it is difficult to either discover or exclude with confidence. This results from the fact that DM and neutrino events can only be distinguished on a statistical level, not event-by-event, and so even if this discrimination of the two populations is clear the limit will scale at best as $\sigma_{\mathrm{lim}} \sim (\textrm{exposure})^{-1/2}$.

However this separation is not always clear, leading to the second effect of the neutrino floor.
This arises for DM masses where the recoil spectrum is indistinguishable from that expected for solar neutrinos. In this case the limiting cross section $\sigma_{\mathrm{lim}}$ improves even slower than $\sim (\textrm{exposure})^{-1/2}$, due to the systematic uncertainties on the size of the solar neutrino flux~\cite{Billard:2013qya,Davis:2014ama}. Indeed $\sigma_{\mathrm{lim}}$ reaches a saturation value set by the size of these uncertainties, at which it stays effectively constant even with increasing exposure.
Assuming that the DM velocity distribution $f(v)$ is a Maxwell-Boltzmann distribution then the strongest effect is for DM masses around 6~GeV, however this extends to heavier masses when the uncertainties in $f(v)$ are taken into account~\cite{Davis:2014ama}. Fortunately this second effect of the neutrino floor exists only when using a single experiment with only information on the spectrum used to separate DM and neutrino recoils. Indeed it is strongly reduced if either a second experiment is used in combination with the first~\cite{Ruppin:2014bra}, or if the different annual modulation~\cite{Davis:2014ama} or directional dependence~\cite{Grothaus:2014hja,O'Hare:2015mda} of the solar neutrinos and DM is employed as an additional discrimination parameter\footnote{An interesting point is that when these additional discrimination parameters take effect $\sigma_{\mathrm{lim}}$ can improve faster than $\sim (\textrm{exposure})^{-1/2}$. However this is only when the limiting cross section is far from the fundamental Poisson limit, which can not be surpassed. Upon approaching this limit the Poisson scaling of $\sigma_{\mathrm{lim}} \sim (\textrm{exposure})^{-1/2}$ is regained.}. 

Examples of experiments for which this neutrino background will be important are the new SuperCDMS at SNOlab~\cite{cdms_sno_pres}, particularly due to its low threshold, and multi-tonne xenon experiments such as LZ~\cite{2011arXiv1110.0103M}, XENONnT~\cite{2012arXiv1206.6288A} and DARWIN~\cite{2012JPhCS.375a2028B}. There is additionally another background from atmospheric and diffuse supernovae neutrinos which will be important for the sensitivity of these experiments to heavier DM, with masses around 15~GeV and above. However since the flux of such backgrounds is much smaller than that from solar neutrinos they will be far less problematic.

\section{Conclusion}
We have reviewed the recent progress which has been made by direct detection experiments in searching for light dark matter, with mass $m$ in the range $3\,\mathrm{GeV} \lesssim m \lesssim 20\,\mathrm{GeV}$, interacting with nucleons. Much of the recent history of this field has been concerned with tension between null results from experiments such as LUX~\cite{Akerib:2013tjd}, SuperCDMS~\cite{Agnese:2014aze} and XENON100~\cite{Aprile:2012nq} and potential signals of dark matter in the CoGeNT~\cite{Aalseth:2012if}, CDMS-Si~\cite{Agnese:2013rvf}, CRESST-II~\cite{Angloher:2011uu} and DAMA/LIBRA~\cite{Bernabei:2013xsa,Bernabei:2008yi,Bernabei:2013cfa} experiments (see figure~\ref{fig:dm_exclusion}). The first three of these `signal' experiments observe additional events (`excesses') above their background predictions, which could be interpreted as nuclear recoils induced by dark matter from the halo of our galaxy. The DAMA experiment instead claims to observe an annual variation in their data consistent with the expectation from dark matter~\cite{Bernabei:2013xsa,Bernabei:2008yi,Bernabei:2013cfa}.

We discussed in section~\ref{sec:past} recent re-analyses of data from CoGeNT and CRESST-II which showed that these excess events are entirely attributable to underestimated backgrounds, and hence there is in fact no significant preference for dark matter recoils in these data~\cite{Davis:2014bla,Angloher:2014myn}. 
For CoGeNT this took the form of an underestimated background from events on the surface of the detector, whose spectrum mimics the low-energy rise expected from light dark matter~\cite{Davis:2014bla}. While for CRESST-II the background originated from secondary products of $^{206}$Pb nuclei scattering near the surface of the clamps holding the detector~\cite{Kuzniak:2012zm}. The spectrum of this background was softer than initially predicted due to the rough surface of the clamps which the original analysis~\cite{Angloher:2011uu} had assumed this to be perfectly smooth~\cite{Kuzniak:2012zm}. New results from a run without these clamps confirm this null result~\cite{Angloher:2014myn}.

In section~\ref{sec:dama} we discussed the interpretations of the data from the DAMA/LIBRA experiment~\cite{Bernabei:2013xsa,Bernabei:2008yi,Bernabei:2013cfa} either in terms of dark matter or a more mundane source. The DAMA experiment observes an annually varying rate of events whose phase is close to that expected from dark matter. However in order to produce such a modulation the dark matter needs to scatter off nuclei with a large cross section. We discussed difficulties in creating a model for dark matter which can provide such a cross section while evading the upper limits from null searches. We also focused on explanations for this modulation which do not invoke dark matter. In this context we discussed models which can give the same modulation as a dark matter signal by combining signals from atmospheric muons and solar neutrinos, whose rates also vary throughout the year with different phases~\cite{Davis:2014cja}. Finally we summarised the three different results which a second DAMA-like experiment in a different location could observe, and what they would imply for the DAMA signal. A second result with the same phase as DAMA would favour dark matter, a result with a different phase favours a mundane explanation such as muons or neutrinos while no modulation at all favours an explanation in terms of a detector effect in DAMA.

The recent progress in the field of light dark matter direct detection can be summarised as such (see also the review of ref.~\citen{Schumann:2015wfa}):
\begin{itemize}
\item The CoGeNT and CRESST-II `excesses' have been fully explained as due to underestimated backgrounds, and not dark matter recoils~\cite{Davis:2014bla,Angloher:2014myn,Kuzniak:2012zm}.
\item The CDMS-Si excess events are still unexplained (at least publicly). However since the SuperCDMS experiment~\cite{Agnese:2014aze} is more sensitive and finds no evidence for excess events above background then the former is unlikely to be due to dark matter.
\item The DAMA annual modulation remains unexplained, however it is unlikely to be due to dark matter considering the strong constraints from null searches. Much progress has been made recently on explaining the data from DAMA without dark matter, and experiments aimed at replicating this experiment such as DM-Ice will hopefully bring this issue to a close soon~\cite{Davis:2014cja,Cline:2015yza}.
\end{itemize}

In section~\ref{sec:nu_bg} we discussed the future of light dark matter direct detection in terms of the background from solar neutrinos. Future multi-tonne direct detection experiments face the prospect of distinguishing a potential dark matter signal from the large and irreducible background due to solar neutrinos. Such a background will fundamentally limit the sensitivity of these experiments to signals larger than the Poisson uncertainties on the neutrino background. Sensitivity for light dark matter can be optimised by combining results from multiple experiments~\cite{Ruppin:2014bra}, or by using the different annual modulation~\cite{Davis:2014ama} or directional dependence~\cite{Grothaus:2014hja,O'Hare:2015mda} of the dark matter and neutrino signals.

The direct detection of light dark matter faces an exciting future with larger and more sensitive experiments currently under construction or in development~\cite{Cushman:2013zza}. This is especially true for light dark matter due to the development of experiments with low energy thresholds such as SuperCDMS~\cite{Agnese:2014aze} and improvements in the understanding of how xenon-based experiments (e.g. XENON100~\cite{Aprile:2012nq} and LUX~\cite{Akerib:2013tjd}) respond to recoils with energies below 3~keV~\cite{lux_leff_pres}. The presence of a background from solar neutrinos will reduce the rate at which future sensitivity improves and more work needs to be done to understand its full effects, however it should never realistically present an absolute limit on this sensitivity.

\section*{Acknowledgments}
The research of the author is supported at IAP by  ERC project 267117 (DARK) hosted by Universit\'e Pierre et Marie Curie - Paris 6.

\appendix

\section{Dark matter scattering in direct detection experiments \label{sec:app1}}
The spectrum of dark matter recoils in a given detector (in units of counts per day per kg per keV)  takes the form of \cite{Cerdeno:2010jj},
\begin{equation}
\frac{\mathrm{d}R}{\mathrm{d}E} = \frac{\rho_{\chi}}{m_N m} \int_{v_{\mathrm{min}}}^{\infty} v f(v + u_e) \frac{\mathrm{d}\sigma}{\mathrm{d}E} \mathrm{d}^3 v ,
\label{eqn:rate1}
\end{equation}
where $m_N$ is the mass of the nucleus in the detector, $m$ is the DM particle mass, $\rho_{\chi}$ is the local DM density, generally taken to be around 0.3 GeVcm$^{-3}$ \cite{Widrow:2008yg}, $v_{\mathrm{min}} = \sqrt{E m_N / 2 \mu_N^2}$, $\mu_N$ is the DM-nucleus reduced mass, and $ \frac{\mathrm{d}\sigma}{\mathrm{d}E}$ is the differential interaction cross section. The velocity integral accounts for the fact that a DM particle does not have to deposit all of its energy in the detector upon collision, and so any particle with a velocity greater than $v_{\mathrm{min}}$ can impart a kinetic energy of $E$ to the nucleus. All velocities in equation \ref{eqn:rate1} are in the Earth's rest frame, hence we use $u_e$ to boost the distribution of galactic DM velocities $f(v)$ into the correct frame. Since the relative direction of the Earth's velocity with respect to the DM wind varies over the year the rate ${\mathrm{d}R} / {\mathrm{d}E}$ exhibits an annual modulation, which is what the DAMA/LIBRA collaboration claim to observe in their data.

Since the dark matter in the galactic halo moves at non-relativistic velocities this formula is simplified by expanding the differential cross section in terms of recoil velocity $v$ and taking only the lowest-order term. This leads to the expression for the spin-independent scattering cross section,
\begin{equation}
 \frac{\mathrm{d}\sigma}{\mathrm{d}E} = \frac{\sigma m_N F(E)}{2 \mu_N^2 v^2},
\label{eqn:rate2}
\end{equation}
where $\sigma$ is the `zero-momentum' DM-nucleus cross section. The function $F(E)$ is the nuclear form-factor, which for spin-independent interactions is essentially a Fourier transform of the nucleus \cite{Lewin199687}. Hence the expression for the DM-nucleus recoil rate simplifies to
\begin{equation}
\frac{\mathrm{d}R}{\mathrm{d}E} = \frac{\sigma \rho_{\chi}  F(E)}{2 \mu_N^2 m} \int_{v_{\mathrm{min}}}^{\infty} \frac{f(v + u_e)}{v}  \mathrm{d}^3 v .
\label{eqn:rate3}
\end{equation}

The final step is to express this in terms of the scattering cross section between dark matter and \emph{nucleons}, which is the cross section all direct detection experiments set upper limits on (or preferred regions for positive results). If we assume that the DM couples equally to protons and neutrons we obtain for the cross section,
\begin{equation}
\sigma(E) = \sigma_0 \left( \frac{\mu_N}{\mu_p} \right)^2 A^2 ,
\label{eqn:rate4}
\end{equation}
where $\sigma_0$ is the zero-momentum DM-nucleon cross section, $\mu_p$ is the DM-proton reduced mass and $A$ is the atomic number of the nucleus with which the DM interacts. Assuming equal couplings to protons and neutrons is not necessary, and relaxing this assumption may reduce or enhance the rate depending on the particular nuclear target as discussed in section~\ref{sec:dama}.  

If instead the dark matter couples to nuclei via their spin then the scattering is said to be `spin-dependent'. In this case instead of equation (\ref{eqn:rate2}) we have for the differential cross section the expression,
\begin{equation}
\frac{\mathrm{d} \sigma}{\mathrm{d} E} = \frac{16 m_N}{\pi v^2} \Lambda^2 G_F^2 J(J+1) \frac{S(E)}{S(0)},
\end{equation}
where $G_F$ is the Fermi constant, $J$ is the total spin of the nucleus, $S(E)$ is the spin form factor and $\Lambda = \frac{1}{J} [a_p \langle S_p \rangle + a_n \langle S_n \rangle]$, with $\langle S_p \rangle$ and $\langle S_n \rangle$ being the expectation values for the spin of the proton and neutron respectively, and $a_p$ and $a_n$ are coupling constants for the proton and neutron.

\bibliographystyle{ws-ijmpa}

\end{document}